# Magnetically tunable Shubnikov-de Hass oscillations in $MnBi_2Te_4$

X. Lei[1,2], L. Zhou[1], Z. Y. Hao[1], H.T. Liu[2], S. Yang[2], H. P. Sun[1], X. Z. Ma[1], C. Ma[1], H. Z. Lu[1,3], J. W. Mei[1,3,*], J. N. Wang[2,†], and H. T. He[1,4,‡]

[1]*Department of Physics, Southern University of Science and Technology, Shenzhen 518055, China*

[2]*Department of Physics, The Hong Kong University of Science and Technology, Clear Water Bay, Hong Kong, China*

[3]*Shenzhen Institute for Quantum Science and Engineering, Southern University of Science and Technology, Shenzhen 518055, China*

[4]*Shenzhen Key Laboratory for Advanced Quantum Functional Materials and Devices, Southern University of Science and Technology, Shenzhen 518055, China*

**Abstract.** Shubnikov-de Hass oscillations are directly observed in undoped antiferromagnetic topological insulator $MnBi_2Te_4$. With increasing magnetic fields, the oscillation period decreases gradually in the magnetic transition from canted antiferromagnetism to ferromagnetism and then saturates in high magnetic fields, indicating the field-induced evolution of the band structure. From the analysis of the high-field oscillations, a nontrivial Berry phase and a small effective mass are extracted, in agreement with the predicted Weyl semimetal phase in ferromagnetic $MnBi_2Te_4$. Furthermore, rotating the magnetization of $MnBi_2Te_4$ can lead to a splitting of the high-field oscillations, which suggests the enhanced asymmetry of the Weyl cones in tilted fields. Therefore, the observation of these magnetically tunable quantum oscillations clearly demonstrates the indispensable role of field in tuning the band structure or physical properties of $MnBi_2Te_4$.

* meijw@sustech.edu.cn
† phjwang@ust.hk
‡ heht@sustech.edu.cn

Antiferromagnetic topological insulator (AFM TI) $MnBi_2Te_4$ (MBT) has attracted intense attention as an ideal platform for studying the interplay between topology and magnetism [1-3]. Depending on its dimension and magnetic state, MBT can exhibit rich exotic topological quantum states [3]. Quantum anomalous Hall effect (QAHE) and the axion insulator phase have been observed experimentally in odd- and even-layer antiferromagnetic MBT thin films, respectively [4, 5]. With all magnetic moments polarized in high magnetic fields, ferromagnetic (FM) MBT thin films can even exhibit QAHE at temperatures up to 45 K [6], much higher than magnetically doped TIs [7]. In its bulk form, MBT is predicted to be an antiferromagnetic TI with gapped topological surface states due to the $\Theta\tau_{1/2}$ symmetry breaking, where $\Theta$ is the time-reversal symmetry and $\tau_{1/2}$ is the half translation operator connecting nearest spin-up and -down Mn atomic layers [1-3]. But it will change into a type-II Weyl semimetal (WSM) with only a pair of Weyl nodes once it is driven into the ferromagnetic state by an external magnetic field [2, 3]. Besides this, tilting the direction of the magnetic field can also have a profound effect on the phase transition of ferromagnetic MBT, such as the one from type-II to type-I WSM [8].

All these studies demonstrate that the topological quantum state or the band structure of MBT is closely associated with its magnetic state and can be effectively tuned via an external magnetic field. Up to now, ARPES has been widely adopted to investigate the band structure of antiferromagnetic MBT [9-12], but it is incapable of studying the evolution of the band structure with increasing magnetic fields. In contrast, the magnetic field induced Shubnikov-de Hass (SdH) oscillations are able to uncover the close connection between the band structure and magnetic state of MBT. Recently, there have been some reports on the observation of SdH oscillations in Sb-doped MBT, with the focus to justify the appearance of type-II WSM phase in ferromagnetic state [13, 14]. The SdH oscillations were only studied in high perpendicular magnetic fields. But in order to reveal the intriguing field-induced evolution of band structures mentioned above, a systematic study of SdH oscillations in a wide field range and even at different tilting angles is highly desired.

In this work, we report on the direct observation of SdH oscillations in high-quality undoped MBT. It's shown that the SdH oscillations begin to appear when the MBT is in the canted AFM state. But with increasing fields, the oscillation period decreases. Only in high fields when the FM state is stabilized, single-period SdH oscillations are obtained. By analyzing these high-field SdH oscillations, a nontrivial Berry phase and a small effective mass can be extracted, consistent with the WSM phase in FM MBT [2, 3, 13, 14]. More importantly, tilting the magnetization direction of ferromagnetic MBT is found to result in a splitting of the SdH oscillations. Such a behavior is believed to arise from the enhanced asymmetry of the Weyl cones in tilted fields. Our work thus reveals the close connection between the band structure of MBT and its magnetic state through the SdH oscillation measurement, demonstrating the effective manipulation of the band structure by external magnetic fields.

The MBT flakes studied in this work were exfoliated from high-quality undoped MBT crystals synthesized by the flux method. For more details about the crystal growth, the device fabrication, and the measurement method, please refer to our previous study of linear magnetoresistance (LMR) in MBT [15]. It's worth pointing out that the carrier density of our samples is at the order of $10^{19}$ $cm^{-3}$, about one order of magnitude lower than the carrier density obtained in Ref. [16, 17]. Fig. 1 (a) shows the temperature ($T$) dependence of resistance ($R$) of our sample, clearly displaying a resistance peak at 23.6 K. It's the magnetic transition from paramagnetism to AFM that gives rise to this resistance peak, and thus the Neel temperature ($T_N$) is about 23.6 K [9-12, 16, 17]. Fig. 1 (b) shows the measured magnetoresistance (MR) and Hall resistance ($R_{xx}$ & $R_{yx}$) at $T$=2 K. From the MR curve, the critical fields at which the field-induced magnetic transitions from AFM to canted AFM and then to FM can be estimated, *i.e.*, $B_{c1}$=3.5 T & $B_{c2}$=8 T, as indicated in Fig. 1 (b). Since stronger SdH oscillations are observed in the $R_{yx}(B)$ curve, we will focus on the study of Hall resistance in the following.

We first subtract a smooth background obtained by adjacent averaging from the original $R_{yx}(B)$ curve and then plot the obtained $\Delta R_{yx}$ as a function of $1/B$. As shown in Fig. 2 (a), the $\Delta R_{yx}(1/B)$ curve exhibits clear oscillations, indicating the occurrence of SdH oscillations in our undoped MBT samples. But different from previous studies of Sb-doped samples where single-frequency oscillations were observed [13,14], the $\Delta R_{yx}(1/B)$ curve in Fig. 2 (a) deviates from periodic oscillations, especially in the low field region. To see this more clearly, the oscillation period $T_n$ is calculated for each oscillation at 2 K with $T_n = \frac{1}{B_n} - \frac{1}{B_{n+1}}$, where $\frac{1}{B_n}$ denotes the $n^{th}$ peak position, as illustrated in Fig. 2 (a). The obtained period $T_n$ is then plotted as a function of $\bar{B}_n$ in Fig. 2 (b), with $\bar{B}_n = (B_n + B_{n+1})/2$. One can see that the oscillation period decreases obviously with increasing fields. A total decrease of about 70% is observed, which apparently cannot be understood in terms of the conventional single-frequency SdH oscillations. Besides this, it is also noticed that the oscillation period tends to saturate in the high field region.

In order to understand these phenomena, we recall that recent theoretical studies have revealed the sensitivity of band structure to the magnetic ordering in MBT [2, 3, 8]. A-type antiferromagnetic MBT is an antiferromagnetic TI, but when all the Mn moments are polarized in high perpendicular fields, ferromagnetic MBT turns into a type-II WSM with only a pair of Weyl nodes [2, 3]. Even the magnetization direction has an important effect on the band structure of FM MBT. A phase transition from type-II to type-I WSM is expected to occur when tilting the magnetization direction of MBT [8]. All these indicate that the band structure of MBT is closely associated with its magnetic state. As mentioned in Fig. 1 (b), the transition field $B_{c2}$ from canted AFM to FM is about 8 T. Therefore, in the field region studied in Fig. 2 (a), the magnetic state of MBT changes gradually from canted AFM to FM. Considering the close connection between the band structure and the magnetic state of MBT, a change in the band structure is thus expected, which is reflected in the change of the oscillation period of SdH oscillations shown in Fig. 2 (b). Furthermore, the saturation of the oscillation period can be understood, since the MBT will finally reach a stable FM state in high fields.

To further support the above scenario, we have also studied the evolution of SdH oscillations with temperatures. Fig. 3 (a) shows the $\Delta R_{yx}(1/B)$ curves extracted at different temperatures. With increasing temperatures, the oscillation amplitudes decrease as expected. But a shift of the oscillation peaks is also noticed. For clarity, the temperature dependence of each peak position is shown in Fig. 3 (b). Increasing temperatures leads to an obvious shift of the low-field peak positions ($n$= 9 & 10). On the contrary, the high-field oscillation peaks ($n$=6, 7, and 8) exhibit little shift. This phenomenon can also be understood in the same scenario. For high-field peaks, the corresponding magnetic state is a fully polarized ferromagnetic state. But the low-field peaks appear in the canted AFM or weak FM states, which are more vulnerable to the temperature. Therefore, with increasing temperatures, the low-field magnetic state will change more than the high-field FM state. Due to the close connection between the band structure and the magnetic state discussed above, one can thus expect a larger change in the low-field oscillations with increasing temperatures, manifested as the shift of the peak positions in Fig. 3 (b). Similar phenomena have also been observed in the study of SdH oscillations in Sb-doped MBT [13, 14].

From the above analysis of SdH oscillations shown in Fig. 2 & 3, it is clear that the band structure of MBT is indeed closely associated with its magnetic states, in agreement with previous theoretical expectations [2, 3, 8]. It also demonstrates that the external magnetic field can be an effective knob in tuning the physical properties of MBT, as well as the temperature [13, 14].

According to the Onsager relation $A_F \frac{\hbar}{eB_n} = 2\pi\left(n + \frac{1}{2} - \frac{\phi_B}{2\pi} - \delta\right) = 2\pi(n+\gamma)$, the Landau level index $n$ is a linear function of $1/B_n$, where $A_F$ is the extremal cross-sectional area of Fermi surface perpendicular to the field, $\hbar$ is the Plank's constant, $e$ is the electronic charge, $B_n$ is the magnetic field where the SdH oscillation peak with Landau level index $n$ is located, $\phi_B$ is a geometrical phase, also known as the Berry phase, and $\delta$ is a phase shift depending on the dimensionality of the system [19]. The slope and $n$-intercept of the $n\left(\frac{1}{B_n}\right)$ curve would yield the values of $\frac{A_F\hbar}{2\pi e}$ and $\gamma$, respectively. In Fig. 4 (a), we plot the $\frac{1}{B_n}$ dependence of $n$ at 2 K, with the peaks (or valleys) of the SdH oscillations indexed by integers (or half-integers) (see Fig. 2 (a)). Note that we only analyze the oscillation data with $B >$ 10 T where stable FM ordering is reached in MBT. As expected, $n$ is linearly dependent on $\frac{1}{B_n}$. From the linear fitting curve shown in Fig. 4 (a), $\gamma = -0.19$ are obtained. For $n$-type Weyl semimetals breaking the time-reversal symmetry, such as ferromagnetic MBT, $\gamma$ is expected to be $-1/8$ if the Fermi energy is right at the Weyl nodes [20]. The small deviation of $\gamma$ from $-1/8$ can be understood since the Fermi energy is above the Weyl nodes in our MBT samples. Besides the nontrivial Berry phase, WSM is expected to have tiny effective mass due to the linear energy dispersion around the Weyl nodes. In order to estimate the effective mass of MBT, we first perform fast Fourier transform (FFT) of the SdH oscillations with $B > 10$ T at different temperatures as shown in Fig. 4 (b) and then plot the FFT amplitudes as a function of $T$ in Fig. 4 (c). It's noted that the obtained FFT frequency $f$ is only about 80 T, indicating that the Fermi energy is rather close to the charge-neutral point of MBT [14]. According to the Lifshitz-Kosevich (LK) formula [19, 21], the temperature dependent SdH oscillation amplitude is described by the thermal damping factor $R_T = [(aTm^*/B)/sinh(aTm^*/B)]$, where $a = 2\pi^2 k_B/(e\hbar)$, $m^*$ is the effective mass, and $k_B$ is the Boltzmann constant. Therefore, by fitting the FFT amplitudes to the damping factor $R_T$, we derive the value of the effective mass $m^* = 0.043m_0$, where $m_0$ is the free electron mass. It's worth pointing out that the obtained $m^*$ is close to the theoretical values calculated for the conduction band of MBT with the Fermi energy close to the Weyl nodes [13, 14]. The presence of a nontrivial Berry phase and a tiny effective mass thus provide evidence for the existence of the WSM phase in FM MBT, consistent with previous SdH studies of Sb-doped MBT [13, 14].

We also studied the SdH oscillations in tilted magnetic fields, as shown in Fig. 5. The tilting angle $\theta$ is defined as the angle between the field vector and the $z$ direction, as illustrated in the inset of Fig. 5. All the curves are plotted as a function of the reciprocal of $Bcos(\theta)$, *i.e.*, the $z$ component of the field. As the field is tilted away from the perpendicular direction ($\theta = 0^o$), the obtained SdH oscillations changes, showing non-2D behaviors. The most prominent feature is the splitting of the $n$ =7 peak when the tilting angle is larger than $10^o$, as indicated by the two dashed lines in Fig. 5. The splitting increases gradually with increasing tilting angles. It's believed that the $n = 6$ peak also splits, as the oscillation amplitude weakens in tilted fields. But the magnetic field range of our equipment (< 16 T) is not large enough for us to show such a splitting in high fields.

As the $n$ =7 peak shows no splitting at $\theta = 0^o$, the isotropic spin splitting is not responsible for the splitting observed with $\theta > 10^o$. It should be the anisotropic orbital splitting that gives rise to the phenomenon. It's noted that although the anisotropic orbital splitting has been reported in previous studies of semimetals such as $Cd_3As_2$ or $ZrTe_5$ [22-25], the splitting decreases with increasing tilting angles. But in our case, this orbital splitting is enhanced as the tilting angles increase. Since the orbital splitting is closely associated with the shape of the band structure, the observed splitting in Fig. 5 indicates that the tilted field induces a change in the band structure of MBT. This is qualitatively in agreement with a recent theoretical study where a phase transition of FM MBT from type-II to type-I WSM is predicted by tilting the magnetic field [8]. It's also noted that such a transition occurs when the tilting angles exceed $10^o$, the same as the critical angle at which the splitting in Fig. 5 begins to appear. This coincidence might suggest a close relationship between these two phenomena, but more efforts are still needed in the near future to clarify this issue.

Based on the above results and discussions, we propose the following scenario to account for the enhanced splitting in tilted fields in FM MBT. In high fields, the ferromagnetic MBT is a type-II WSM with only a pair of Weyl nodes [2, 3]. In our MBT sample with the carrier concentration at the order of $10^{19}$ cm$^{-3}$, the Fermi level should lie above the Weyl nodes. The electrons from each Weyl cone should contribute to the observed SdH oscillations in Fig. 5. Since no splitting is present at $\theta = 0^o$, it is argued that the asymmetry between the two Weyl cones is negligible, *i.e.* the Fermi pocket is almost the same for the two Weyl cones. As mentioned before, tilting the field will lead to a change in the band structure of ferromagnetic MBT [8]. The emergence of splitting with $\theta > 10^o$ in Fig. 5 thus indicates that the asymmetry between the two Weyl cones is enhanced with tilting angles, or the difference in the Fermi pocket from each Weyl cone becomes more prominent at higher tilting angles. It is this enhanced asymmetry in tilted high fields that leads to the splitting of the $n$ =7 peak in Fig. 5. It's noted that the above scenario is proposed with the assumption that the Fermi level is below the Lifshitz point of MBT. If the Fermi level is above the Lifshitz point, a single Fermi pocket instead of two will be expected. Considering the non-degenerate property of the band, the single Fermi pocket cannot give rise to the splitting of SdH oscillation peak. The assumption is justified since the density functional theory calculations of the band structure of MBT have shown that MBT with $f$ ~ 80 T has a Fermi energy below the Lifshitz transition point [14]. Therefore, the splitting shown in Fig. 5 not only provides more evidence for the existence of the WSM phase in high fields, but also reveals the evolution of the two Weyl cones with the tilting angles of the field. It's worth pointing out that asymmetric WSMs have been predicted to exhibit some intriguing phenomena, such as the giant photocurrent and the chiral magnetic effect without chirality source [26, 27].

In conclusion, we have directly observed SdH oscillations in undoped MBT, which shows sensitively dependence on the magnetic states of MBT. Single-frequency oscillations can only be obtained when the fully polarized FM state is stabilized in high fields. The nontrivial Berry phase and the tiny effective mass extracted from the analysis of the SdH oscillations strongly indicate the existence of the WSM phase in FM MBT. Furthermore, rotating the direction of magnetization in tilted fields can lead to a splitting of the SdH oscillation peak, suggesting the enhanced asymmetry of the Weyl cones in MBT. Therefore, our study clearly demonstrates magnetically tunable SdH oscillations in MBT and reveals the important role of magnetization or field in tuning the band structure of MBT.

## ACKNOWLEDGEMENTS

This work was supported by the National Key Research and Development Program of China (No. 2016YFA0301703), the Science, Technology and Innovation Commission of Shenzhen Municipality (No. GXWD20201230110313001 & No. ZDSYS20190902092905285), the Natural Science Foundation of Guangdong Province (No. 2021A1515010046), and the Research Grant Council of Hong Kong SAR (No. 16301418 & No. C6025-19GF).

X. Lei and L. Zhou contributed equally to this work.

## FIGURE CAPTIONS

FIG 1. (a) The temperature dependent resistance of MBT flakes. Inset: An enlarged view of the $R(T)$ curve enclosed by the dashed rectangle. (b) The longitudinal and Hall resistance obtained in perpendicular magnetic fields at $T$=2 K. Field-induced magnetic transitions from AFM to canted AFM and then to FM occur consecutively at $B_{c1}$ and $B_{c2}$, respectively.

FIG 2. (a) The extracted oscillatory $\Delta R_{yx}(1/B)$ curve obtained at $T$=2 K, with peaks and valleys indexed by integers and half-integers, respectively. (b) The magnetic field dependence of the oscillation period.

FIG 3. (a) The $\Delta R_{yx}(1/B)$ curves obtained at different temperatures. (b) The temperature dependence of each peak positions.

FIG 4. (a) The Landau fan diagram of the SdH oscillations at $T$=2 K. (b) The FFT spectra of the high-field SdH oscillations at different temperatures. (c) The temperature dependence of the FFT amplitude, fitted by the damping factor of the Lifshitz-Kosevich (LK) formula.

FIG 5. The extracted oscillatory $\Delta R_{yx}$ at different tilting angles, plotted as a function of the reciprocal of $Bcos(\theta)$. Inset: $\theta$ is defined as the tilting angle of $B$ with respect to the $z$ direction.

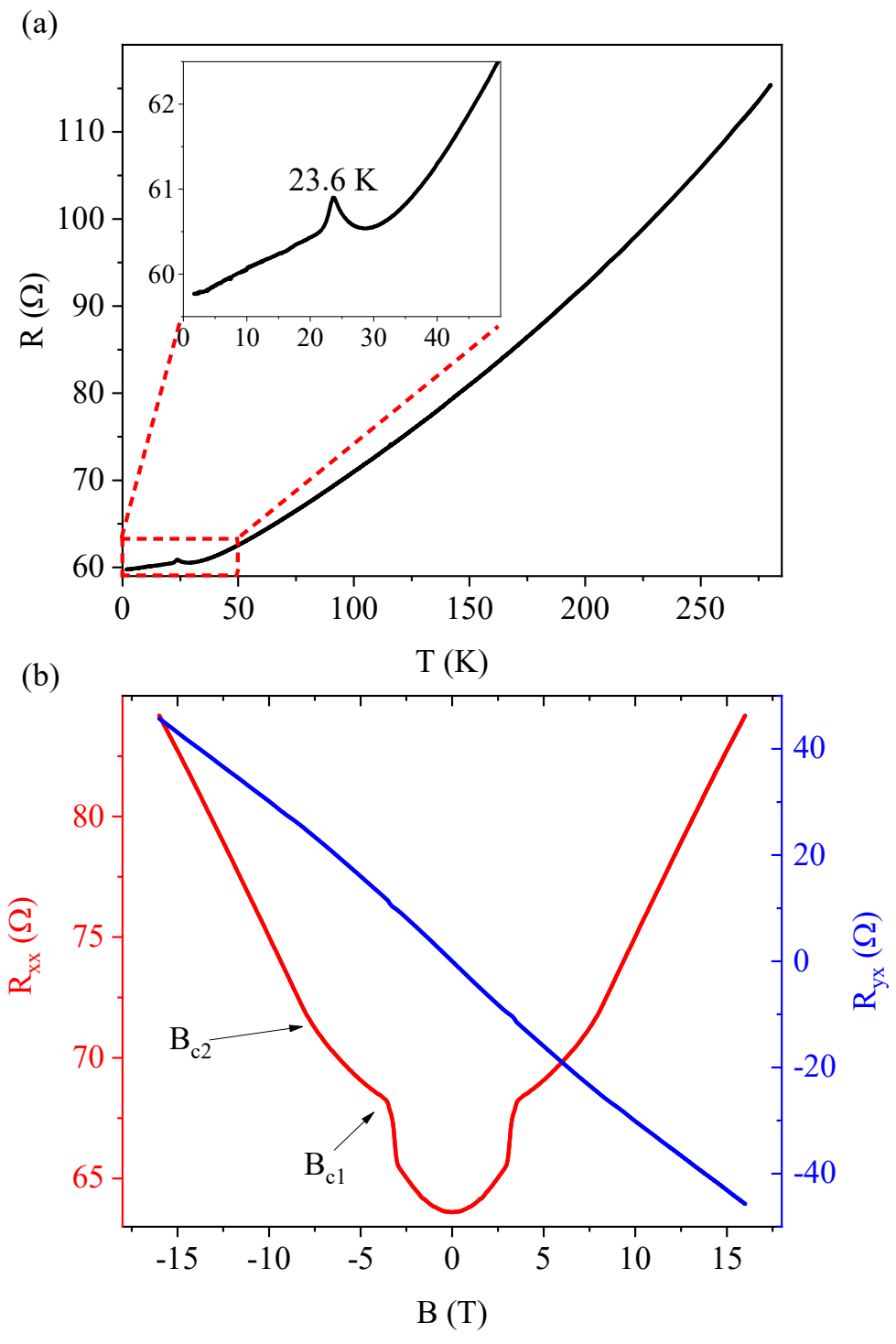

Fig. 1

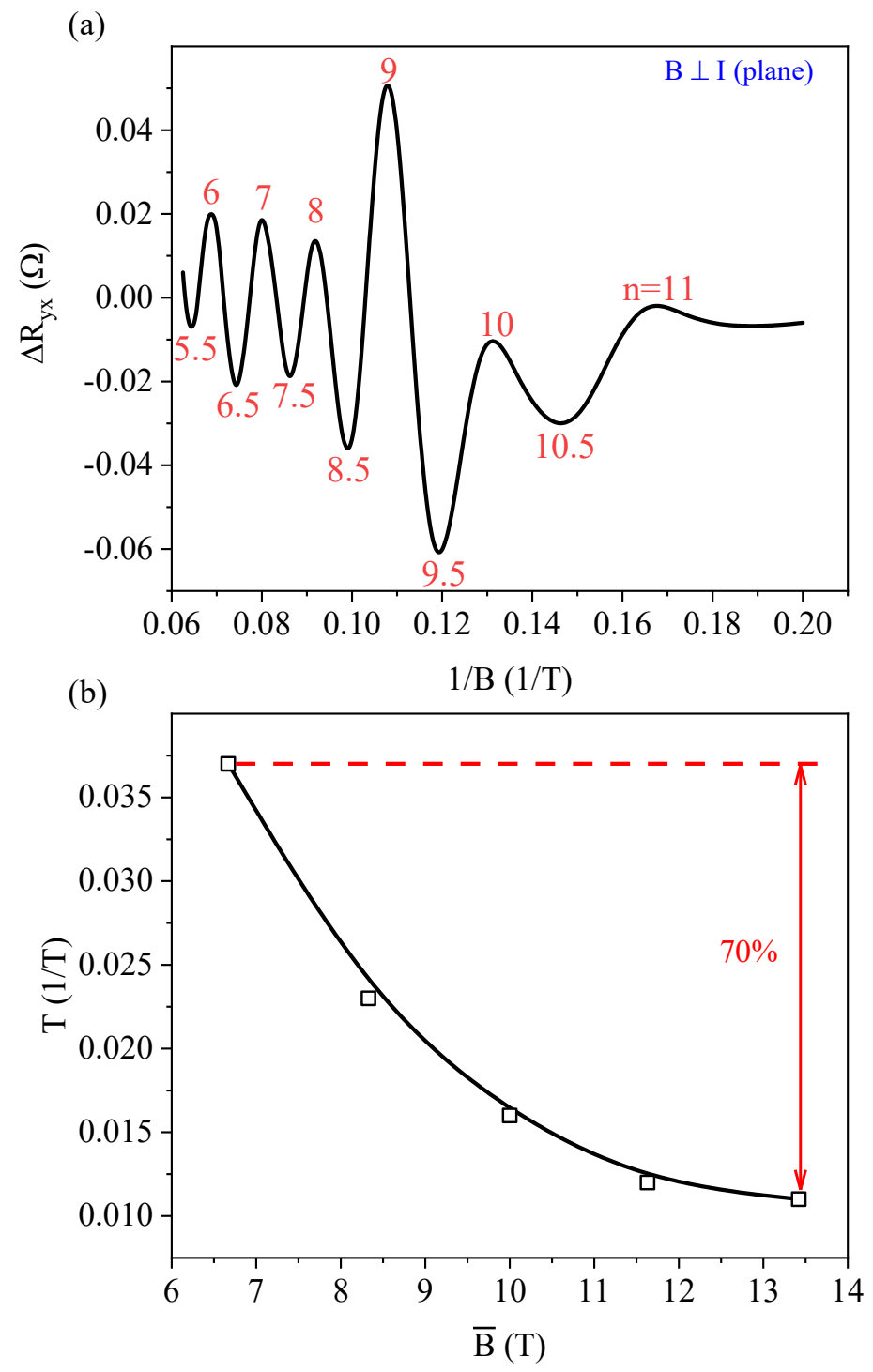

Fig. 2

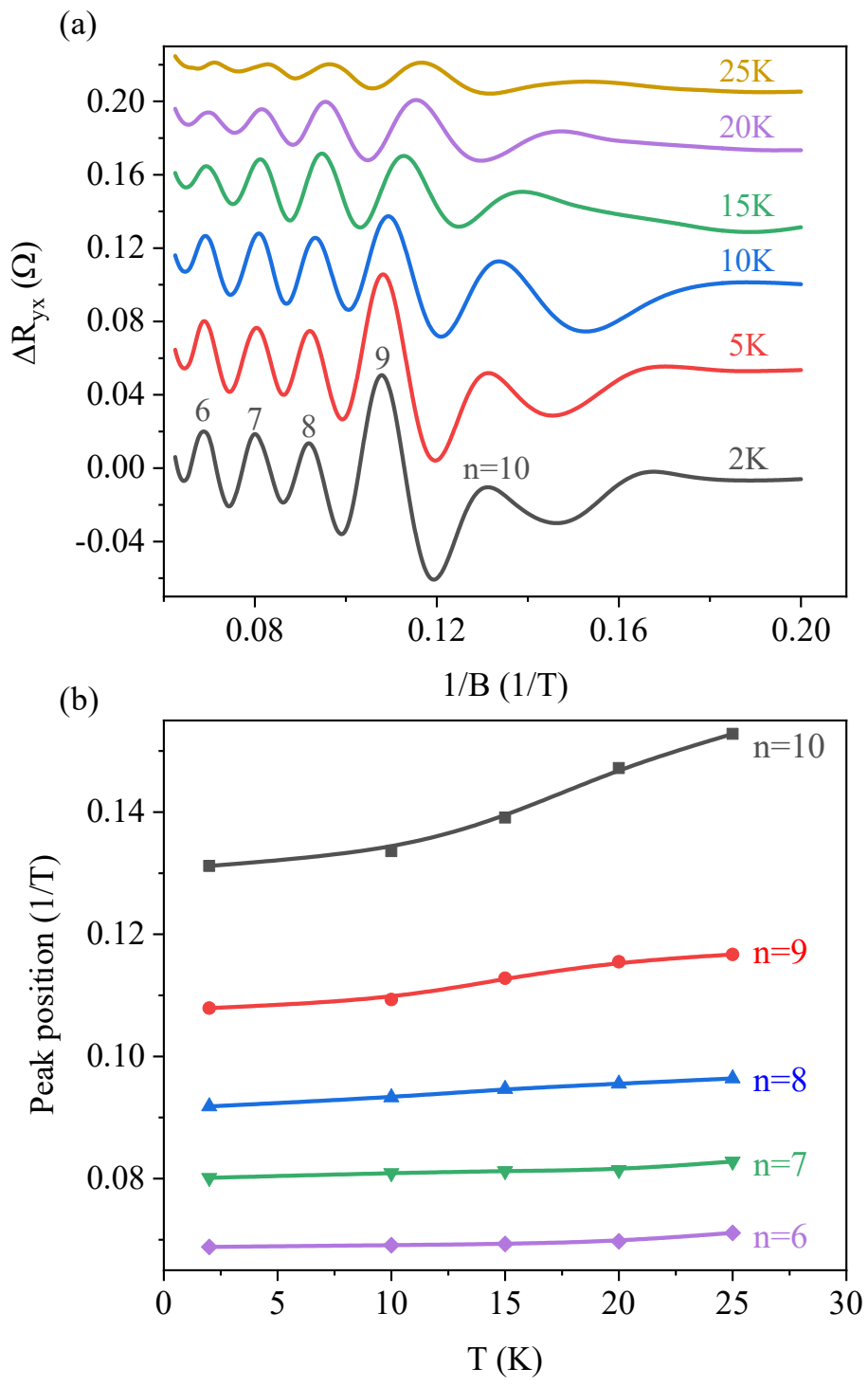

Fig. 3

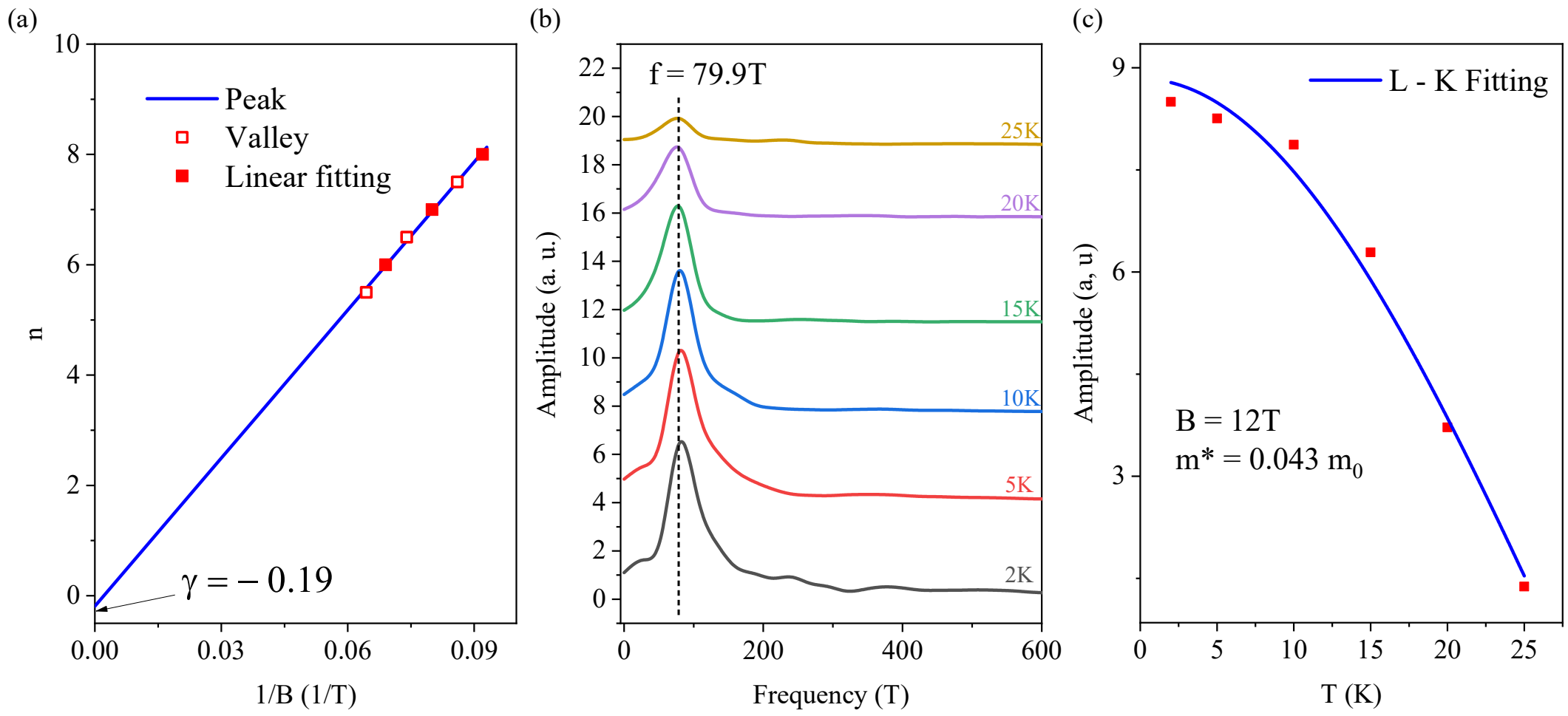

Fig. 4

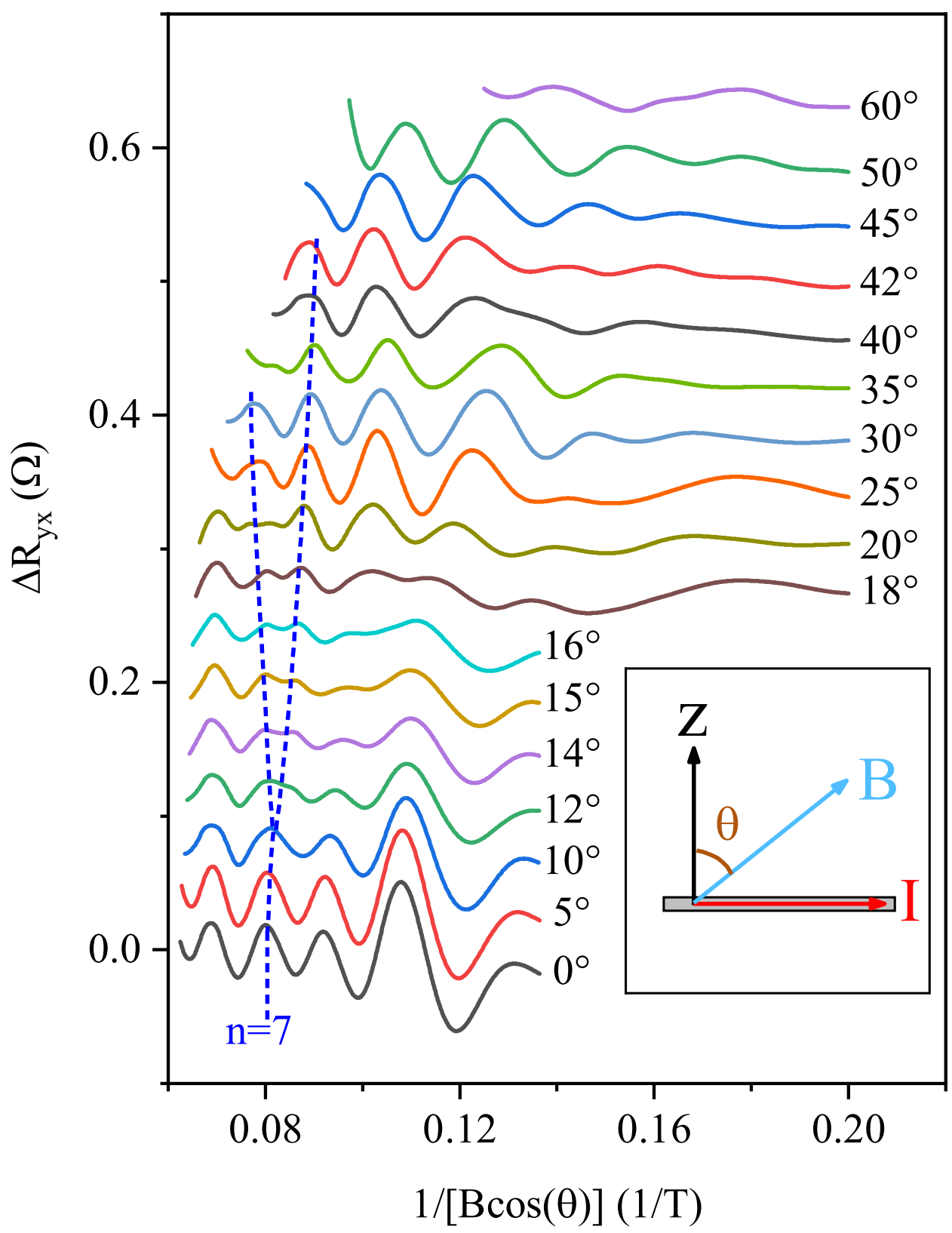

Fig. 5